\documentclass[aps,prl,twocolumn,superscriptaddress,showpacs]{revtex4}
\usepackage{amsfonts}
\usepackage{amssymb}
\usepackage{amsmath}
\usepackage{epsfig}
\usepackage{graphicx}
\usepackage{dcolumn}
\usepackage{bm}
\usepackage{flafter}
\usepackage{float}

\begin{document}

\title{Spin-Chiral Bulk Fermi Surfaces of BiTeI Proven by Quantum Oscillations}
\author{Joonbum Park}
\affiliation{Department of Physics, Pohang University of Science and Technology, Pohang 790-784, Korea}
\author{E. Kampert}
\affiliation{Dresden High Magnetic Field Laboratory, Helmholtz-Zentrum Dresden-Rossendorf, Dresden, D-01314, Germany}
\author{Kyung-Hwan Jin}
\affiliation{Department of Physics, Pohang University of Science and Technology, Pohang 790-784, Korea}
\author{Man Jin Eom}
\affiliation{Department of Physics, Pohang University of Science and Technology, Pohang 790-784, Korea}
\author{Jongmok Ok}
\affiliation{Department of Physics, Pohang University of Science and Technology, Pohang 790-784, Korea}
\author{E. S. Choi}
\affiliation{National High Magnetic Field Laboratory, Florida State University, Tallahassee, Florida 32310, USA}
\author{F. Wolff-Fabris}
\affiliation{Dresden High Magnetic Field Laboratory, Helmholtz-Zentrum Dresden-Rossendorf, Dresden, D-01314, Germany}
\affiliation{European XFEL GmbH, Albert-Einstein-Ring 19, Hamburg, 22761, Germany}
\author{K. D. Lee}
\affiliation{Department of Physics, Inha University, Incheon, 402-201, Korea}
\author{N. Hur}
\affiliation{Department of Physics, Inha University, Incheon, 402-201, Korea}
\author{J.-S Rhyee}
\affiliation{Department of Applied Physics, Kyung Hee University, Yongin, 446-701, Korea}
\author{Y. J. Jo}
\affiliation{Department of Physics, Kyungpook National University, Daegu 702-701, Korea}
\author{Seung-Hoon Jhi}
\affiliation{Department of Physics, Pohang University of Science and Technology, Pohang 790-784, Korea}
\author{Jun Sung Kim}
\affiliation{Department of Physics, Pohang University of Science and Technology, Pohang 790-784, Korea}
\date{\today}

\begin{abstract}
We present the Fermi-surface map of the spin-chiral bulk states for the non-centrosymmetric semiconductor BiTeI using de Haas-van Alphen and Shubnikov-de Haas oscillations. We identify two distinct Fermi surfaces with a unique spindle-torus-type topology and the non-trivial Berry phases, confirming the spin chirality with oppositely circulating spin-texture. Near the quantum limit at high magnetic fields, we find a substantial Zeeman effect with an effective $g$-factor of $\sim$ 60 for the Rashba-split Fermi surfaces. These findings provide clear evidence of strong Rashba and Zeeman coupling in the bulk states of BiTeI, suggesting that BiTeI is a good platform hosting the spin-polarized chiral states.

\end{abstract}


\maketitle
Exotic electronic orders in crystalline solids arise when the spin degeneracy is lifted by the inversion or time-reversal symmetry breaking. For example, in superconductivity, strong Rashba coupling induces the parity mixing of the spin-singlet and the spin-triplet pairing~\cite{SC:Rashba}, and strong Zeeman coupling produces the spatially-modulated superconducting order known as the Fulde-Ferrell-Larkin-Ovchinnikov phase~\cite{SC:FFLO}. Also, both strong Rashba and Zeeman coupling combined with the superconductivity lead to spin polarized superconductors that can host the Majorana states~\cite{SC:Majorana}. However, experimental realization of these exotic phases in three dimensional (3D) materials, which are good platforms for the electronic orders, is usually limited by very small Rashba and Zeeman coupling strengths. In this respect, the non-centrosymmetric semiconductor BiTeI has attracted considerable interest recently because it can provide an intrinsic 3D bulk system with giant Rashba effect~\cite{BTI:Cal1, BTI:ARPES1, BTI:ARPES2, BTI:ARPES3, BTI:ARPES4, BTI:Optics}.

BiTeI consists of positively charged BiTe layers and negatively charged I layers stacked alternately without the inversion symmetry. \emph{Ab-initio} calculations predicted that the Rashba effect is greatly enhanced for bulk states of BiTeI
with a large Rashba parameter $\alpha_R$ of $\sim$ 3.5 eV/$\rm{\AA}$~\cite{BTI:Cal1}. Experimental verification of the spin-chiral bulk states in BiTeI is essential
and several angle resolved photoemission spectroscopy (ARPES) measurements~\cite{BTI:ARPES1, BTI:ARPES2, BTI:ARPES3, BTI:ARPES4} have been
performed for this purpose. These studies, however, demonstrated that two dimensional (2D) Rashba-split surface states coexist with 3D Fermi surfaces.
This blocks a direct access to the Rashba-split \emph{bulk} Fermi surfaces in BiTeI, and their detailed nature of the spin chirality, the Fermi surface topology, and the strengths of Rashba and Zeeman coupling has not been clarified yet.

In contrast to ARPES, magnetic quantum oscillations at high magnetic fields provide a more sensitive probe for investigating the Fermi surfaces in all crystallographic axes~\cite{Shoenberg} as well as their spin chirality with nontrivial Berry's phases~\cite{Berry:PKim, Berry:Ong, Berry:Ando,Berry:PPhillips,gfactor:Bi2Se3}.
In fact, the quantum oscillations of the resistivity, $i.e.$, the Shubnikov-de Haas (SdH) effect, were employed recently to study the Fermi surfaces of BiTeI~\cite{BTI:QO1, BTI:QO2}. However, the SdH measurements alone cannot completely rule out the possibility of detecting the Rashba-split surface states rather than the bulk states. This contrasts to the de Haas-van Alphen (dHvA) effect, $i.e.$, the quantum oscillations of magnetic susceptibility, which is bulk sensitive. In this Letter, we present the full map of the bulk Fermi surfaces of BiTeI obtained from both dHvA and SdH measurements over wide ranges of magnitude and tilting angle of magnetic fields. We unambiguously determine the spindle-torus-type topology, the oppositely circulating spin chirality, and the significantly large $g$-factor in the Rashba-split bulk Fermi surfaces of BiTeI.
These results suggest that BiTeI provides a promising platform to incorporate spin-polarized bulk states, which can host intriguing electronic orders without spin degeneracy.

\begin{figure}
\includegraphics*[width=8.5cm, bb=10 15 480 420]{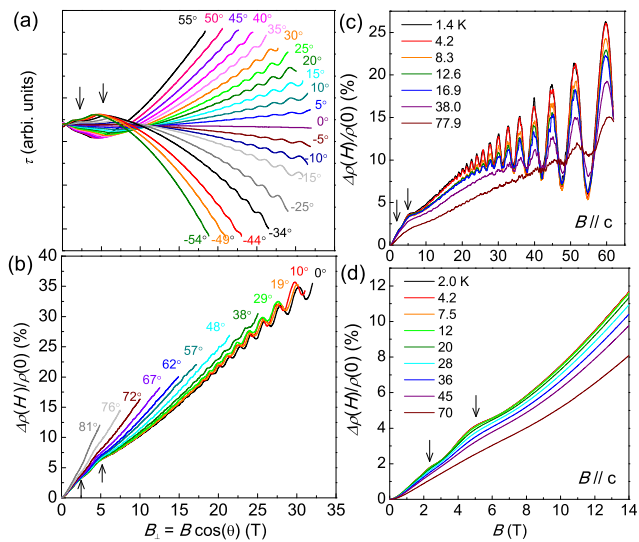}
\caption{\label{Fig.1} (color online) (a) Torque signal $\tau$ (S1) and (b) the in-plane resistivity $\rho$ of BiTeI (S2) as a function of normal magnetic fields, $B_{\perp}$ $=$ $B\cos\theta$ under static magnetic fields up to 32 T at $T$ = 0.35 K for different angle $\theta$ with respect to the $c$-axis. The in-plane resistivity of BiTeI at different temperatures (c) under pulsed magnetic fields up to 62 T (S3) and (d) under static magnetic fields of up to 14 T (S4).}
\end{figure}

Single crystals of BiTeI were grown using the Bridgman method~\cite{BTI:ARPES1}. For the dHvA experiments, we employed torque magnetometry. Small single crystals with a size of about 30$\times$30$\times$10 $\mu$m$^3$ were mounted on the miniature Seiko piezo-resistive cantilever. To study the SdH effects, the in-plane resistivity was measured in the standard 6-probe geometry. In total, four samples (S1$\sim$S4) from the same batch were measured in magnetic fields up to 14 T using a Physical Properties Measurement System (Quantum Design), to 33 T using a Bitter magnet at the National High Magnetic Field Lab. (NHMFL), Tallahassee, USA, and to 62 T using a pulsed field magnet at the Hochfeld-Magnetlabor (HLD), Dresden, Germany. First principles calculations were carried out in the plane-wave basis within the local-density and generalized gradient approximations (LDA and GGA) for exchange-correlation functionals~\cite{perdew,perdew2}. We used the Vienna \emph{ab-initio} simulation package~\cite{kresse}. A Cutoff energy of 400 eV was chosen in the plane-wave basis set to expand the wave-functions and atomic potentials. Employing the experimental lattice parameters $a$ $=$ 4.340 $\rm{\AA}$ and $c$ = 6.854 $\rm{\AA}$~\cite{shevelkov}, the ionic positions were fully optimized until the forces on each ion became less than 0.01 eV/$\rm{\AA}$. The k-point integration was done by sampling the Brillouin zone with $k$-point meshes of 10$\times$10$\times$8 grids.

Figure 1(a) shows the magnetic field dependence of the torque signal $\tau$ as the magnetic field is rotated from $B$ $\parallel$ [001] ($\theta$ = 0$^{\rm o}$) towards $B$ $\parallel$ [100] ($\theta$ = 90$^{\rm o}$) at $T$ = 0.35 K. $\tau(B)$ tends to curve with a dip or a hump near $B$ $\sim$ 5 T, followed by the $B^2$-dependence at higher magnetic fields, which is in contrast to the conventional $B^2$ dependence. Since the torque magnetometry measures the anisotropy of magnetic susceptibility ($\Delta\chi = \chi_c-\chi_{ab}$) as described by $\tau$ $\propto$ $B^2\Delta\chi\sin(2\theta)$~\cite{note_tau}, initial decrease of $\tau (B)$ in our measurement should be related to a strong diamagnetic signal for BiTeI. In fact, a recent study on the magnetic susceptibility of BiTeI demonstrated a large diamagnetic susceptibility ($\chi_c$ $<$ 0) while keeping a paramagnetic $\chi_{ab}$ when the Fermi level is near the band crossing point~\cite{BTI:orbital}. In addition to the background signal in $\tau(B)$, clear dHvA oscillations were observed [Fig. 1(a)]. We found slow oscillations at relatively low magnetic fields (denoted by the arrows in Fig. 1) and fast oscillations at higher magnetic fields above $B$ $\sim$ 11 T~\cite{note_dHvA}. The SdH oscillations in the in-plane resistivity $\rho$ upon variation in angle $\theta$ [Fig. 1(b)] and temperature [Fig. 1(c)-(d)] also show consistent results with Fig. 1(a).

Unlike the SdH effect, the dHvA effect has dominant sensitivity on the bulk states over the surface states. The dHvA effect may be found in 2D electron systems, producing small saw-tooth-type oscillations due to the chemical potential rearrangement~\cite{dHvA:2DEG}. But, this is not the case in our results. The consistent results obtained from dHvA and SdH measurements provide direct evidence of monitoring the bulk Fermi surfaces of BiTeI.
The dHvA and SdH oscillations of BiTeI differ essentially from the typical behaviors of Rashba-split Fermi surfaces in semiconductor heterostructures. For a small Rashba splitting, the inner and outer Fermi surfaces should have similar sizes, and produce a beating pattern in the quantum oscillations as found in $e.g.$, InGaAs/InAlAs heterostructures~\cite{Rashba:transport}. In BiTeI, however, we observed two well-separated frequencies.
For $\theta$ $\approx$ 0, frequencies of $F\sim$4.5(6)T and $F\sim$360(10)T correspond to the inner and the outer Fermi surfaces, respectively. From the Onsager relation $F$ = ($\Phi$$_0$/2$\pi^{2}$)$S_F$, where $\Phi$$_0$ is the flux quantum and $S_F$ the Fermi surface size, we estimate $S_F$ to be 0.043(5) nm$^{-2}$ and 3.48(7) nm$^{-2}$ for each oscillation, consistent with recent SdH measurements~\cite{BTI:QO1, BTI:QO2}. Such a large difference in the size of Rashba-split Fermi surfaces confirms the giant bulk Rashba effect in BiTeI.

\begin{figure}
\begin{center}
\includegraphics*[width=8.0cm, bb=0 215 580 800]{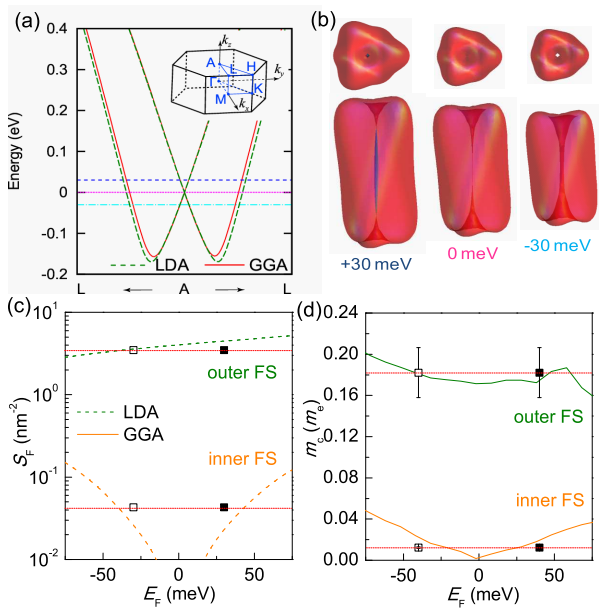}
\caption{\label{FIG. 2} (color online) (a) Band dispersion along the symmetry line $L$-$A$-$L$. The Brillouin zone of BiTeI is given in the inset. (b) The Fermi surfaces for $E_F$ = 30, 0, and -30 meV, which are indicated as dashed lines in (a). (c) The size ($S_F$) and (d) the cyclotron mass ($m_c$) as a function of $E_F$. The dashed (solid) lines indicate the results using LDA (GGA) functionals. The experimentally determined $S_F$ and $m_c$ are also plotted with (red) solid lines.}
\end{center}
\end{figure}

Having established that the quantum oscillations originate from the 3D bulk states, now we discuss the Fermi surface topology of BiTeI. As the Fermi level $E_F$ crosses the degenerate point ($E_F$= 0), the bulk Fermi surface of BiTeI undergoes a topological transition from a ring-torus-type ($E_F$ $<$ 0) to a spindle-torus-type ($E_F$ $>$ 0) [Fig. 2(a) and (b)]. Accompanied by the topology change, the spin textures of the inner and the outer Fermi surfaces are also significantly modified. In the Rashba-split Fermi surfaces, the circulating direction of the polarized spins is opposite in the upper and the lower bands. Therefore, for the spindle-torus-type ($E_F$ $>$ 0), electron spins in the inner and the outer Fermi surfaces circulate in the opposite direction. In contrast, for the ring-torus-type ($E_F$ $<$ 0), they circulate in the same direction.

The experimental determination of the Fermi surface topology, however, is hampered by subtle differences in the shapes of the spindle-torus-type and the ring-torus-type Fermi surfaces. For example, the intensity map of bulk-sensitive soft X-ray ARPES cannot distinguish them due to insufficient resolution~\cite{BTI:ARPES3,BTI:ARPES4}. Comparing the Fermi surface sizes ($S_F$) and the cyclotron mass ($m_c$) from SdH measurements with \emph{ab-initio} calculations~\cite{BTI:QO2} is also not enough either to determine the Fermi surface topology. In Fig. 2(c) and (d), we show the calculated $S_F$ and $m_c$ for the inner and the outer Fermi surfaces as a function of $E_F$, using LDA and GGA functionals. We note that the sizes of the Fermi surfaces~\cite{supp} are slightly larger in the LDA case than in the GGA case [Fig. 2(a)]. The experimental values of $S_F$ and $m_c$ for the inner and the outer Fermi surfaces are in reasonable agreement with the LDA calculations by assuming $E_F$ $\approx$ $-$30 meV and the ring-torus-type Fermi surfaces~[Fig.~2(c) and (d)]. However, the comparison with the GGA calculations yields $E_F$ $\approx$ 30 meV and the spindle-torus-type Fermi surfaces.

\begin{figure}
\begin{center}
\includegraphics*[width=8.0cm, bb=30 20 570 570]{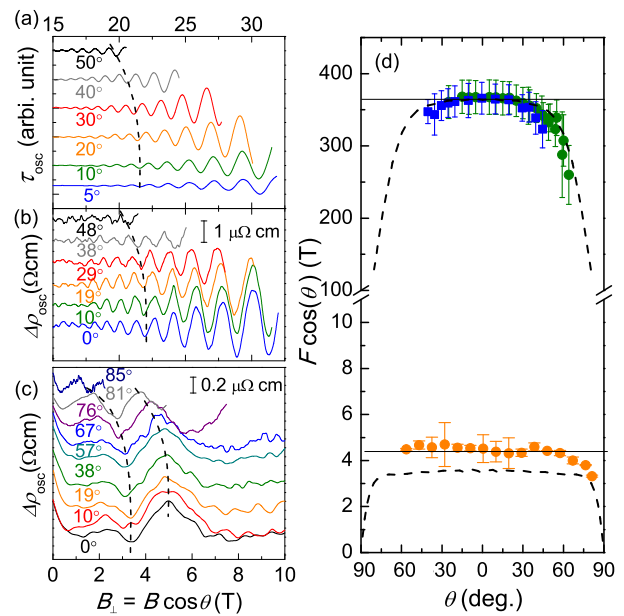}
\caption{\label{FIG.3} (color online)(a) The oscillatory part of the torque signal ($\tau_{osc}$) for S1 and (b),(c) the resistivity ($\Delta\rho_{osc}$) for S2 as a function of normal magnetic fields $B_{\perp}$ $=$ $B\cos\theta$ at 0.35 K at different tilting angles $\theta$. The dashed lines are guides to the eye. (d) Angular dependence of observed dHvA and SdH frequencies ($F\cos\theta$) as a function of $\theta$ for S1 (squares) and S2 (circles). The dashed lines are the results from the \emph{ab-initio} calculations with GGA.}
\end{center}
\end{figure}
The \emph{angle dependence} of the dHvA and SdH oscillations unambiguously determines the Fermi surface topology. Figure 3 shows the oscillatory part of the torque signal ($\tau_{osc}$) and the resistivity ($\Delta\rho_{osc}$) as a function of the out-of-plane magnetic field $B_{\perp}$ = $B$cos($\theta$). At low angles, the oscillations show a scaling of the peak positions with $B_{\perp}$, while such a scaling breaks down at higher angles~[Fig. 3(a)$-$(c)]. The corresponding frequencies~\cite{note_FFT} with the tilted angles $\theta$ follow $F$ $\propto$ 1/cos($\theta$) at low $\theta$'s for both the inner and the outer Fermi surfaces. Such a quasi-2D behaviour is consistent with the previous studies~\cite{BTI:QO2}. The clear deviation from the 2D behaviour is only observed at $\theta$ $>$ 70$^{\rm o}$ for the inner Fermi surface and at $\theta$ $>$ 40$^{\rm o}$ for the outer Fermi surface. These results clearly demonstrate that SdH and dHvA measurements over a wide range of tilting angles are essential for revealing the highly-elongated 3D Fermi surfaces of BiTeI.

 A key difference between the ring-torus-type and the spindle-torus-type Fermi surfaces is observed in the curvature of the inner Fermi surface along $k_z$. For the ring-torus-type, the cross-section increases more rapidly with the tilting angle $\theta$ than the 2D cylindrical case. But, for the spindle-torus-type, it decreases rapidly with $\theta$ at high angles. Thus the angle dependence of the inner Fermi surface close to the in-plane magnetic field is critical for determining the Fermi surface topology of BiTeI. The $F$$\cos \theta$ for the inner Fermi surface [Fig.~3(c)] suddenly decreases at $\theta$ $>$ 70$^{\rm o}$, consistent with the spindle-torus-type FS. The calculated $F\cos\theta$ assuming $E_F$ = 30 meV in the GGA calculations shows a good agreement with the experiments. This confirms the Rashha energy of $E_{R}$ $\sim$ 150 meV for the bulk Fermi surfaces of BiTeI [Fig. 3(a)]. We note that the Rashba energy $E_{R}$ is much larger than the Fermi energy of $E_F$ $\approx$ 30 meV. Hence, our system is close to the transition of the Fermi surface topology, which can be induced by tuning the $E_F$ across the degenerate point with slight hole doping~\cite{BTI:ARPES3}.

\begin{figure}
\begin{center}
\includegraphics*[width=8.5cm, bb=30 450 550 730]{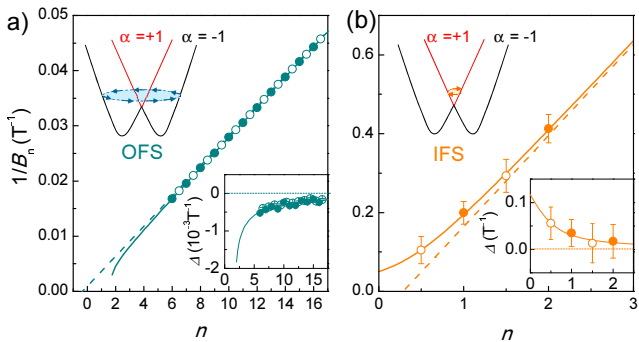}
\caption{\label{FIG. 4} (color online) The Landau fan diagram for (a) the outer and (b) the inner Fermi surfaces for S3. The schematic dispersion of Rashba split bands ($\alpha$, the band index) and the corresponding Fermi surfaces are shown in the inset. The solid lines are the fitting curves taking into account the Zeeman effect (see the text), and the dashed lines are the asymptotic curves without the Zeeman effects. The insets in (a) and (b) show a deviation of the experimental data from the asymptotic curve without the Zeeman effects.}
\end{center}
\end{figure}

Now we discuss the spin chirality and the Zeeman effect in the quantum oscillations of BiTeI. For the Rashba-split Fermi surfaces, the electron momentum-spin locking produces non-trivial Berry's phase in the cyclotron orbit, which gives an additional phase offset ($\gamma$) in the quantum oscillations~\cite{Berry:PPhillips, Berry:Mikitik,gfactor:Bi2Se3, Berry:mackenzie}. The phase offset is estimated from the so-called Landau fan diagram, where the extrema of the quantum oscillations ($1/B_n$) are plotted with the Landau index ($n$) as shown in Fig. 4. For a bulk system that has the resistivity ($\rho_{xx}$) much larger than the Hall resistivity ($\rho_{xy}$), the integer filling factor $n$ is assigned to the maximum of the $\rho_{xx}$ because $\Delta \sigma_{xx}$ $\sim$ $-\Delta \rho_{xx}$ ~\cite{Berry:PPhillips}. In this case, the extrapolation of the Landau fan diagram to the limit of $1/B$ $\rightarrow$ 0 determines the phase offset $\gamma$. For the spindle-torus-type Fermi surfaces, the spins in the inner Fermi surface circulate in the opposite direction to those of the outer Fermi surface. Then, $\gamma$ is expected to be $\pm$0.5 corresponding to the non-trivial Berry's phase of $\pm\pi$ for the inner ($\alpha$ = $+$1) and the outer ($\alpha$ = $-$1) Fermi surfaces. The Landau fan diagram of BiTeI, however, exhibits a clear deviation from the linear dependence at higher magnetic fields~[Fig. 4]. This leads to $\gamma$ $\sim$ $-$0.1 for both Fermi surfaces, far from the expected values of $\pm$0.5. Similar behaviour was also reported for topological insulators such as Bi$_2$Se$_3$ ~\cite{gfactor:Bi2Se3}.

According to recent theoretical calculations~\cite{Berry:mackenzie,gfactor:Ando,Berry:Mikitik}, the phase offset $\gamma$ depends on the magnetic field, because the Zeeman gap is induced near the quantum limit at high magnetic fields. In this case, the inner Fermi surface shrinks whereas the outer Fermi surface expands with increasing magnetic fields. This leads to nonuniform Landau level spacing at high magnetic fields, which is indeed found in our experiments. As shown in the inset of Fig.~4, the deviation from the linear dependence shows positive (negative) curvature for the inner (outer) Fermi surfaces. Then $\gamma$ near the quantum limit of high magnetic fields deviates from the zero-field limit of the phase offset $\gamma_{B\rightarrow0}$~=~$\pm$0.5 which corresponds to the intrinsic Berry's phase~\cite{Berry:Mikitik}. Therefore, the significant nonlinear dependence of the Landau fan diagram indicates strong Zeeman effects in BiTeI with a large effective $g$-factor.

The zero-field phase offset ($\gamma_{B\rightarrow0}$) and the effective $g$-factor are estimated using the procedure of Wright and Mckenzie~\cite{Berry:mackenzie}. As shown in Fig. 4, the Landau fan diagram for both the Fermi surfaces is fitted quite well by the equation of \begin{math}n = \frac{F}{B} - \gamma_{B\rightarrow0} - C[\frac{d \gamma}{dB}]_{B\rightarrow0} B \end{math}. Here, $C[\frac{d \gamma}{dB}]_{B\rightarrow0}$ is the parameter describing the effects of the Zeeman gap and the electron-hole asymmetry~\cite{Berry:mackenzie}. The fitting yields the zero-field phase offset of $\gamma_{B\rightarrow0}$~=~ 0.3(2) and -0.35(9) for the inner and the outer Fermi surfaces, respectively. While a slight deviation from the ideal value of $\pm$0.5 is observed, which is presumably due to the warping of the Fermi surface~\cite{Shoenberg}, the non-trivial $\gamma_{B\rightarrow0}$ confirms the spin-chirality in the bulk Fermi surfaces of BiTeI. In particular, the opposite signs of $\gamma_{B\rightarrow0}$ for the inner and the outer Fermi surfaces reveal the oppositely circulating direction of their polarized spins. This is clearly consistent with the spindle-torus-type topology of the Fermi surface as found in Fig. 3.

Effective $g$-factor from the fitting is estimated to be $g$ $\sim$ 56 and $g$ $\sim$ 63 for the inner and the outer Fermi surfaces, respectively~\cite{note_g}. These values are significantly larger than the typical $g$-factor of $\sim$ 2, but similarly large $g$-factor was found in Bi-based narrow-gap semiconductors such as Bi$_2$Se$_3$ ~\cite{gfactor:bulk}. Typical conditions for large $g$-factor in semiconductors are known to be a small energy gap, strong spin-orbit interaction, and symmetry of the energy bands~\cite{g-factor:cal}. These conditions are well-satisfied in the electronic structures of BiTeI, which are in fact considered as the origin of the giant Rashba effect~\cite{BTI:Cal1}.
With a large $g$-factor of $\sim$ 60, the Zeeman energy can be comparable to the Fermi energy of $E_F$~$\approx$ 30~meV at a magnetic field of $B$ $\sim$ 10 T. We note that the Rashba energy $E_R$ $\sim$ 150~meV in BiTeI is also well above the Fermi energy $E_F$. When the Rashba energy and Zeeman energy exceed the Fermi energy, the spin-polarized chiral states are realized as found in the surface states of topological insulators. This implies that BiTeI at moderate magnetic fields or with exchange fields near the thin film of ferromagnetic metals can support the spin-polarized chiral states, providing a good platform of realizing exotic electronic orders without the spin degeneracy.

In summary, based on dHvA and SdH oscillations of BiTeI single crystals, we clarified the detailed nature of the bulk Fermi surfaces of BiTeI: the spindle-torus-type topology, the oppositely circulating spin chirality, and the significantly large Zeeman effects with $g$-factor of $\sim$ 60. Exploiting the 3D nature of BiTeI, we expect that various phase transitions can be explored by modifying its electronic structure with $e.g.$, pressure or chemical doping. For example, recent theoretical calculations predicted that external pressure induces the band inversion, which converts BiTeI into topological insulators~\cite{BTI:Cal2}. Enhanced superconducting instability was also proposed due to the effective reduction of the dimensionality in the low-density region of the Rashba system~\cite{SC:Rashba2}. Strong Rashha and Zeeman coupling, therefore, render BiTeI as a promising material for efficient spin control using electric or magnetic fields as well as for hosting various topological quantum phase transitions.

\begin{acknowledgments}
Authors thank Y. W. Son for helpful discussions. This work was supported by the Mid-Career Researcher Program (No. 2012-013838), SRC (No. 2011-0030785), the Max Planck POSTECH/KOREA Research Initiative Program (No. 2011-0031558) by the National Research Foundation of Korea (NRF) funded by the Ministry of Education, Science and Technology. The work at the HLD was supported by EuroMagNET II under the EU contract number 228043. The work at the NHMFL was supported by National Science Foundation Cooperative Agreement No. DMR-1157490, the State of Florida, and the U.S. Department of Energy.

\end{acknowledgments}


\begin{thebibliography}{00}
\bibitem{note_tau} The torque magnetometry measures the aniostropy of the magnetic susceptibility ($\Delta\chi$ = $\chi_c-\chi_{ab}$) as described by $\tau$ $=$ $\textbf{m}\times\textbf{B}$ $=$ $\frac{1}{2}$$\mu_0$$V$$H^2$$\Delta\chi$sin2$\theta$, where $\mu_0$ and $V$ are the vacuum permeability and the sample volume, respectively.
\bibitem{note_dHvA} Due to the nonmonotonous background in $\tau(B)$ at low magnetic fields, the detailed analysis for the smaller orbit corresponding to the inner FS was carried out mostly for the SdH results.
\bibitem{note_FFT} The high frequency for the outer Fermi surface is estimated by the fast Fourier transform of the oscillations, while the low frequency for the inner Fermi surface is estimated by the slope of the Landau fan diagram.
\bibitem{note_g} For the detailed information, see Eqs. (23)-(27) and (36)-(40) in Ref.[\onlinecite{Berry:mackenzie}]. Here we used the calculated values of $m$ = 0.1 $m_e$ and $v_F$ = 5.83$\times10^{5}$ m/s for describing the electron-hole asymmetry in BiTeI.
\end{thebibliography}
\end{document}